\documentclass[twocolumn,times,final]{elsarticle}

\usepackage{prletters}
\usepackage{framed,multirow}

\usepackage{amssymb}
\usepackage{latexsym}

\usepackage{amsmath}
\usepackage{siunitx}
\renewrobustcmd{\bfseries}{\fontseries{b}\selectfont}
\newrobustcmd{\B}{\bfseries}
\usepackage{booktabs}
\usepackage{multirow}
\usepackage{url}
\usepackage{subfigure}
\usepackage{lscape}
\usepackage{color}

\usepackage{url}
\usepackage{xcolor}
\definecolor{newcolor}{rgb}{.8,.349,.1}

\journal{Pattern Recognition Letters}

\begin{document}

\setcounter{page}{1}

\begin{frontmatter}

\title{Hierarchical Matrix Factorization for Interpretable Collaborative Filtering}

\author[1]{Kai \surname{Sugahara}}\ead{ksugahara@uec.ac.jp}
\author[1]{Kazushi \surname{Okamoto}\corref{cor1}}
\cortext[cor1]{Corresponding author}
\ead{kazushi@uec.ac.jp}

%% \affiliation[label1]{organization={},%Department and Organization
%%             addressline={},
%%             city={},
%%             citysep={}, % use if no comma needed between city and postcode%%             
%%             postcode={},
%%             state={},
%%             country={}}

\affiliation[1]{organization={Department of Informatics, Graduate School of Informatics and Engineering, The University of Electro-Communications},
                addressline={1-5-1 Chofugaoka}, 
                city={Chofu}, 
                postcode={182-8585}, 
                state={Tokyo},
                country={Japan}}

% \affiliation[2]{organization={Affiliation 22},
%                 addressline={Address}, 
%                 city={City}, 
%                 citysep={},
%                 postcode={Postal Code}, 
%                 state={State},
%                 country={Country}}

\received{1 May 2013}
\finalform{10 May 2013}
\accepted{13 May 2013}
\availableonline{15 May 2013}
\communicated{S. Sarkar}

\begin{abstract}
Matrix factorization (MF) is a simple collaborative filtering technique that achieves superior recommendation accuracy by decomposing the user-item interaction matrix into user and item latent matrices.
Because the model typically learns each interaction independently, it may overlook the underlying shared dependencies between users and items, resulting in less stable and interpretable recommendations.
Based on these insights, we propose ``Hierarchical Matrix Factorization'' (HMF), which incorporates clustering concepts to capture the hierarchy, where leaf nodes and other nodes correspond to users/items and clusters, respectively.
Central to our approach, called hierarchical embeddings, is the additional decomposition of the latent matrices (embeddings) into probabilistic connection matrices, which link the hierarchy, and a root cluster latent matrix.
The embeddings are differentiable, allowing simultaneous learning of interactions and clustering using a single gradient descent method.
Furthermore, the obtained cluster-specific interactions naturally summarize user-item interactions and provide interpretability.
{Experimental results on ratings and ranking predictions show that HMF outperforms existing MF methods, in particular achieving a 1.37 point improvement in RMSE for sparse interactions.}
Additionally, it was confirmed that the clustering integration of HMF has the potential for faster learning convergence and mitigation of overfitting compared to MF, and also provides interpretability through a cluster-centered case study.
\end{abstract}

\begin{keyword}
% \MSC 41A05\sep 41A10\sep 65D05\sep 65D17
% \KWD Keyword1 \sep Keyword2 \sep Keyword3
\KWD recommender systems\sep collaborative filtering\sep hierarchical matrix factorization\sep interpretable recommendation

%% MSC codes here, in the form: \MSC code \sep code
%% or \MSC[2008] code \sep code (2000 is the default)
\end{keyword}

\end{frontmatter}

% \linenumbers

%% main text
\section{Introduction}

{A personalized recommendation system suggests individual item sets to users based on their preferences, past interactions, and demographics.
For instance, on a content platform, it can recommend movies or music tailored to the user's habits, significantly enhancing their satisfaction with the service.}
Matrix factorization (MF)~\cite{2009aug_y.koren} is a successful model-based approach for personalized recommender systems.
The main idea is to decompose the matrix representing the observed interactions between users and items into two types of matrices: user-latent and item-latent matrices.
The interaction between a user and item is then predicted by computing the inner product of the obtained user and item latent vectors.
Recognized for its simplicity and power, MF is used in a variety of applications{~\cite{2021nov_y.chen}} and has been applied to a wide range of deep-learning-based methods to improve accuracy~\cite{2017apr_x.he, 2017aug_h.xue}.
On the other hand, understanding the meaning of each dimension of the learned embeddings is usually challenging, hindering its practicality in a recommender system, even though they preserve valuable information for prediction.
Addressing the interpretability or explainability of the model is crucial to improve indicators beyond recommendation accuracy, including factors such as user satisfaction~\cite{2020xxx_y.zhang}.

The shortest way to provide an interpretation of the recommendation model is to generate some reasons by auxiliary use of side information about users and items, such as attribute information and review text, but this content information is usually difficult to obtain, making this a somewhat impractical setting.
Therefore, in limited settings where only user-item interaction matrices are available, interpretation is provided indirectly by presenting users and items that are similar to the recommendation results on the matrix.
As an approach, matrix factorization methods with simultaneous and explicit clustering have been proposed to summarize the users and items in the system, taking into account unobserved interactions.
However, they work on specific prediction tasks (e.g., rating prediction~\cite{2021may_f.almutairi, 2019may_h.li, 2018jun_s.wang} and ranking prediction~\cite{2022nov_a.melchiorre}) and on matrix factorization algorithms designed specifically for them (e.g., non-negative MF).
Therefore, these methods are not directly applicable to newer MF applications, and opportunities for interpretation are lost.

Based on the above insights, we propose a hierarchical matrix factorization (HMF) method that can simultaneously perform prediction and clustering in a single model.
HMF is designed to stably predict unobserved interactions while extracting the hierarchical relationships between users and items in order to provide an abstract interpretation such as ``this group of users strongly prefers this group of items.''
To this end, we further decompose the traditional latent matrix into (a) probabilistic connection matrices representing the hierarchical relationships between objects (i.e., users and items in this study) and clusters, and (b) a latent matrix of root clusters.
Inspired by the motivation behind fuzzy clustering~\cite{2001jul_c.oh}, each object and cluster is represented by a weighted average of the abstract embeddings of its parent clusters.
This simple formulation, called {\em hierarchical embedding}, makes the loss function designed for HMF differentiable as well as conventional MFs and easily extendable to an advanced MF method based on gradient descent.
To evaluate the effectiveness of the proposed method, we conducted two experiments based on rating prediction and ranking prediction.
A comprehensive comparison with vanilla MF and existing hierarchical MF methods verified whether HMF can improve the recommendation accuracy and also learn interactions stably.
In addition, we conducted case studies on a real movie rating dataset and observed the interpretations produced by HMF.

In summary, our main contributions are as follows:
\begin{enumerate}
    \item We propose an end-to-end matrix factorization method HMF that simultaneously detects the user and item hierarchies.
    \item We introduce the {\em hierarchical embeddings}, which are general and differentiable, into MF, allowing HMF to be optimized with a single gradient descent method.
    \item We provide a summary of the cluster-level interactions from the obtained hierarchical embeddings, resulting in the interpretability of HMF.
\end{enumerate}
\section{Related Work}

\subsection{MF and Its Extensions}

MF is often explained in the context of an explicit feedback task, in which users assign a rating to certain items, such as five stars.
Here, we assume $M$ users and $N$ items are available for the recommendation task.
Subsequently, we can construct the rating matrix $\boldsymbol{X} \in \mathbb{R}^{M \times N}$ from the historical interactions, where each element $\boldsymbol{X}_{ij}$ denotes the rating of item $j$ provided by user $i$.
The aim is to model each feedback interaction between user $i$ and item $j$ using the inner product of the $d$-dimensional user and item latent vectors $\boldsymbol{U}_i, \boldsymbol{V}_j \in \mathbb{R}^{d}$; that is, $\boldsymbol{X}_{ij} \approx \boldsymbol{U}_i^T \boldsymbol{V}_j$ for all users and items in the task.
In general, each ``factor'' in a user's latent vector represents a preference of the user, such as liking a certain film genre.
Similarly, each factor in an item's latent vector represents its characteristics.
Thus, computing the inner product allows us to express that a strong match between a user's preferences and an item's characteristics will result in a strong interaction (i.e., a high score) between the user and item.

Not all interactions between users and items can be observed in a real-world setting.
The simplest way to deal with this problem is to treat the unobservables as zeros; however, because the resulting rating matrix is usually very sparse, this leads to an overfitted prediction.
Therefore, the latent vectors are learned from the observed interactions using the objective function:
\begin{equation}
    \min_{\boldsymbol{U}, \boldsymbol{V}} \sum_{(i, j) \in \boldsymbol{\Omega}} \left( \boldsymbol{X}_{ij} - \boldsymbol{U}_i^T \boldsymbol{V}_j \right)^2 + \lambda_{\boldsymbol{\Theta}} ||\boldsymbol{\Theta}||_2^2 \label{eq:mf-objective}
\end{equation}
where $\boldsymbol{\Omega}$ denotes the set of observed interactions,
$\lambda_{\boldsymbol{\Theta}}$ is the regularization parameter, and $\boldsymbol{\Theta}$ represents the set of regularized parameters.
Stochastic gradient descent (SGD)~\cite{2009aug_y.koren} and alternating least squares (ALS)~\cite{2008dec_r.pan} are primarily used as the optimizers.
Then, in the context of the recommender system, the resulting latent matrices are generally used to predict and recommend items that are likely to be highly valued by the user.

There are also many collaborative filtering settings for implicit feedback, such as ``clicks,'' and ``purchases.''
It is usually inappropriate to use Eq. (\ref{eq:mf-objective}) as a binary classification of whether or not there was an interaction, because it overfits for unobserved interactions.
Therefore, pointwise loss~\cite{2017apr_i.bayer, 2008dec_y.hu, 2011dec_x.ning}, pairwise loss~\cite{2009jun.s.rendle}, and softmax loss~\cite{2022xxx_f.ricci} have been proposed instead.
In particular, bayesian personalized ranking (BPR)~\cite{2009jun.s.rendle} is a well-known method owing to its simplicity.
The underlying idea is the probability that an observed interaction $(i, j)$ ranks higher than an unobserved interaction $(i, k)$.
Therefore, the objective function is formulated as follows:
\begin{equation}
    \min_{\boldsymbol{U}, \boldsymbol{V}} - \sum_{\substack{(i, j) \in \boldsymbol{\Omega}, (i, k) \notin \boldsymbol{\Omega}}}
    \ln \sigma(\boldsymbol{U}_i^T \boldsymbol{V}_j - \boldsymbol{U}_i^T \boldsymbol{V}_k)
     + \lambda_{\boldsymbol{\Theta}} ||\boldsymbol{\Theta}||_2^2 \label{eq:bpr-objective}
\end{equation}
where $\sigma$ is a sigmoid function $\sigma(x) = (1 + e^{-x})^{-1}$.
The objective function is generally optimized using SGD and negative sampling (i.e., the sampling of unobserved interactions).

Although MF alone is a powerful tool, several studies have attempted to further improve recommendation accuracy by introducing a neural network architecture~\cite{2017apr_x.he, 2017aug_h.xue}.
The common motivation of them is to overcome the linear model of MF and capture more complex user-item relationships by introducing nonlinearity and allowing representation learning.
Neural matrix factorization (NeuMF)~\cite{2017apr_x.he} combines a generalization of MF with a neural architecture and a multilayer perceptron that captures the nonlinear relationship between users and items.
As these methods can be optimized with SGD-based schemes, the applicability of SGD can easily extend the MF architecture.

\subsection{MF with Clustering}

A pioneering approach is to use the hierarchical structure of objects obtained a priori to resolve sparsity in the factorization~\cite{2012xxx_ali.mashhoori, 2011aug_a.menon}, but in practice such hierarchical information is difficult to obtain.
In addition, Hidden Group Matrix Factorization (HGMF)~\cite{2014nov_x.wang} detects groups of objects in the factorized matrix, but the clustering and decomposition processes are independent, and the classification information obtained during decomposition may be overlooked.
Therefore, such collaborative filtering settings are beyond the scope of our study.

Instead, we focus on frameworks that simultaneously perform matrix factorization and clustering, which is related to several existing studies.
Capturing implicit hierarchical structures for recommender systems (known as IHSR)~\cite{2015jul_s.wang,2018jun_s.wang} was the first approach to further decompose the latent matrices into some non-negative matrices by applying a non-negative MF scheme.
Hidden Hierarchical MF~\cite{2019may_h.li} also tackles hierarchical clustering and rating prediction simultaneously and consists of a bottom-up phase for learning the hidden hierarchical structure and a top-down phase for prediction using a quasi-Newton method.
Learning tree-structured embeddings (known as eTREE)~\cite{2021may_f.almutairi} introduce a new regularization term, which represents the difference between the embedding of each child item and its parent item in a hierarchical structure, and also was optimized in the non-negative MF setup and used for rating prediction.
In addition, the prototype-based MF method should also be mentioned, although it does not explicitly address clustering.
ProtoMF~\cite{2022nov_a.melchiorre} defines a vector that measures the similarity between the embeddings of users and items and the embeddings of prototypes (similar to clusters) to model implicit feedback.
Overall, these methods are limited to specific tasks (i.e., explicit or implicit feedback) and optimization schemes.
It is questionable whether they can be directly applied to MF-based deep-learning methods, which have been increasingly developed in the context of representation learning in recent decades.

\subsection{Interpretable Recommendation}

It is essential to provide users and system vendors with the reasons for recommendations to improve non-accuracy metrics, including the transparency, persuasiveness, and reliability of the recommendations~\cite{2020xxx_y.zhang}.
These techniques (not limited to recommender systems) have been discussed in terms of interpretation and explanation~\cite{2018feb_g.montavon}.
Xian et al.~\cite{2021sep_y.xian} proposed an attribute-aware recommendation that provides attributes corresponding to recommendations, and McAuley et al.~\cite{2013oct_j.mcauley} attempted to explain latent dimensions in ratings by exploiting hidden topics in the review data.
{Returning to the discussion of pure collaborative filtering tasks, there are limited situations where information about users and items, including review text, is available.
In other words, we must provide a useful interpretation using only the available user-item interactions and model structure.
A traditional approach to interpretability in MF is to present a neighborhood-style reason, such as ``your neighbor users rate this item highly,'' which contains the idea of a memory-based approach~\cite{2017aug_b.abdollahi, 2019jul_w.cheng}.
However, because this type of ``reason'' depends on a per-user or per-item basis, it can be difficult to summarize how the model learns the entire dataset as the number of users and items increases.
ProtoMF~\cite{2022nov_a.melchiorre} models user and item prototypes, enabling interpretation of relationships between user prototypes and users, and between item prototypes and items. However, it does not facilitate higher-level interpretations such as understanding the relationships between user and item prototypes.
Instead, by assuming hierarchical clusters of users and items on on the same latent space, our goal is to achieve multiple levels of abstraction in interpretation such as "What group of items does a group of users prefer?"}
\section{Methodology}

\subsection{Representation of Users and Items}

\begin{figure}[t!]
    \centering
    \subfigure[3-layer HMF]{
        \includegraphics[width=8.75cm]{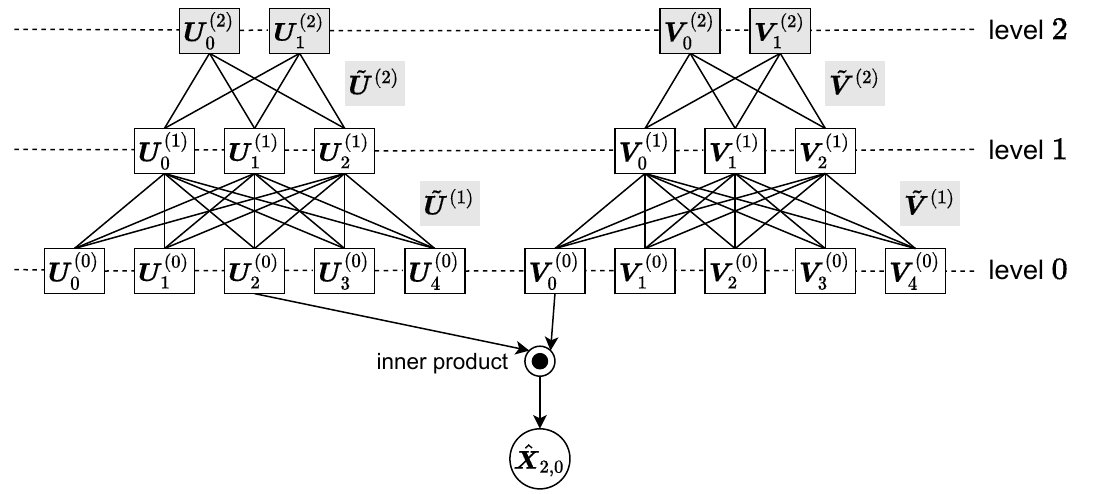}
    }
    \subfigure[MF]{
        \includegraphics[width=8.75cm]{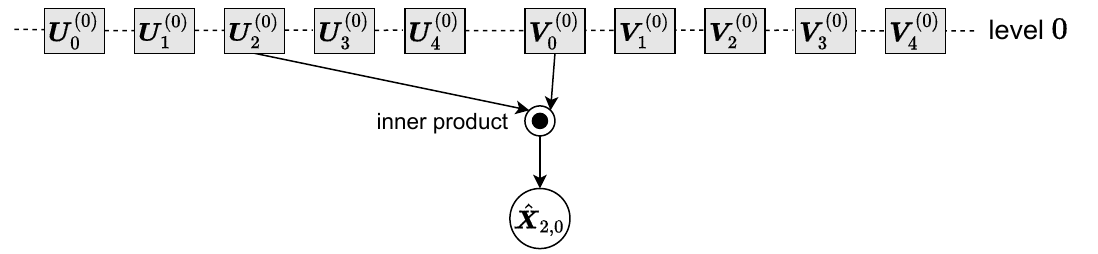}
    }
    \caption{The differences between HMF and MF model architectures illustrated by an example of rating prediction for User 2 and Item 0. Gray highlights indicate the model parameters trained on a dataset. MF has the latent vectors for each user and item, whereas HMF has latent variables in the root clusters and the connection parameters between levels.}
    \label{fig:dif-model-arc}
\end{figure}

This study defines a hierarchical structure with leaf nodes as users/items and other nodes as clusters that abstract them.
A hierarchical structure with a depth of one is a simple clustering setup, whereas a deeper structure captures more complex structures, i.e. clusters of clusters.
With respect to items, it is intuitive to assume a deep hierarchical structure, given that on an e-commerce site, items can be categorized into main categories (e.g., household goods), subcategories (e.g., detergents), and so on.
Here, we assume that each node (i.e., users, items, or non-root clusters) is characterized by its parent clusters; in other words, represented by a more abstract combination of preferences or characteristics.
Consequently, our focus is not on learning the individual user- and item-specific embeddings, but on learning the root cluster embeddings along with their connections (weights) to parent-child nodes, as shown in Fig. \ref{fig:dif-model-arc} (a).
In this study, we refer to the user and item embeddings generated by this scheme as {\em hierarchical embeddings}.
This approach differs from the vanilla MF method, which focuses on learning embeddings that are directly linked to individual objects, as shown in Fig. \ref{fig:dif-model-arc} (b).

For simplicity, we first consider a user hierarchy of depth one (i.e., a non-hierarchical clustering setting).
Let $m_1$ be the number of user clusters; then, we decompose the user latent matrix $\boldsymbol{U}^{(0)} \in \mathbb{R}^{M \times d}$ into the connection matrix $\tilde{\boldsymbol{U}}^{(1)} \in \mathbb{R}^{M \times m_1}$ from users to clusters and the $m_1$ root cluster latent matrix $\boldsymbol{U}^{(1)} \in \mathbb{R}^{m_1 \times d}$; that is, $\boldsymbol{U}^{(0)} = \tilde{\boldsymbol{U}}^{(1)} \boldsymbol{U}^{(1)}$.
We use the constraint $\sum_{s} \tilde{\boldsymbol{U}}_{is}^{(1)} = 1$ for each user $i$, such that the connection matrix represents the probability that users are associated with clusters.
Consequently, each user's embedding (i.e., each row of $\boldsymbol{U}^{(0)}$) is represented by a weighted average of the cluster embeddings.
This is clearly different from IHSR~\cite{2015jul_s.wang,2018jun_s.wang}, which imposes that all elements in its matrices are non-negative real numbers.

In the hierarchical setting, more coarse clusters must be generated from the clusters.
Here, we consider a user hierarchy of depth $p$; that is, we iteratively cluster users or the last generated user clusters $p$ times.
Let the level-specific number of user clusters be $\{m_1, m_2, \dots, m_p\}$ and $m_0 = M$.
This means that there are $m_0$ users at level zero and $m_l$ user clusters at level $l$.
Then, $\boldsymbol{U}^{(0)}$ can be reformulated using recursive decomposition:
\begin{align}
    \boldsymbol{U}^{(0)} &= \tilde{\boldsymbol{U}}^{(1)} \boldsymbol{U}^{(1)} = \tilde{\boldsymbol{U}}^{(1)} \tilde{\boldsymbol{U}}^{(2)} \boldsymbol{U}^{(2)} = \tilde{\boldsymbol{U}}^{(1)} \tilde{\boldsymbol{U}}^{(2)} \cdots \tilde{\boldsymbol{U}}^{(p)} \boldsymbol{U}^{(p)}
\end{align}
where $\tilde{\boldsymbol{U}}^{(l)} \in \mathbb{R}^{m_{l-1} \times m_l}$ represents the connection matrix from user clusters (or user objects) at level $(l-1)$ to user clusters at level $l$, such that $\sum_{s} \tilde{\boldsymbol{U}}_{is}^{(l)} = 1, \forall{i}$, and $\boldsymbol{U}^{(l)} \in \mathbb{R}^{m_l \times d}$ represents the embeddings of user clusters at level $l$.

The item latent matrix can be formulated similarly.
Considering an item hierarchy with depth $q$, let $\{n_1, n_2, \dots, n_q\}$ be the level-specific number of item clusters and $n_0 = N$.
The item latent matrix $\boldsymbol{V}^{(0)} \in \mathbb{R}^{N \times d}$ is expressed recursively as follows:
\begin{align}
    \boldsymbol{V}^{(0)} &= \tilde{\boldsymbol{V}}^{(1)} \boldsymbol{V}^{(1)} = \tilde{\boldsymbol{V}}^{(1)} \tilde{\boldsymbol{V}}^{(2)} \cdots \tilde{\boldsymbol{V}}^{(q)} \boldsymbol{V}^{(q)}
\end{align}
where $\tilde{\boldsymbol{V}}^{(l)} \in \mathbb{R}^{n_{l-1} \times n_l}$ represents the connection matrix from items/item-clusters at level $l-1$ to item clusters at level $l$ and $\boldsymbol{V}^{(l)} \in \mathbb{R}^{n_l \times d}$ represents the embeddings of item clusters at level $l$.
Similarly, the connection matrices $\tilde{\boldsymbol{V}}^{(l)}$ satisfy $\sum_t \tilde{\boldsymbol{V}}_{jt}^{(l)} = 1, \forall{j}$.

Note that these hierarchical embeddings are clearly differentiable with respect to their connection matrices and the latent factors of the top-level clusters.
Thus, the embedding part of the MF, which is differentiable with respect to the latent factors, can be easily replaced by a hierarchical embedding, and SGD can be used for optimization without modification.

\subsection{Loss Function}

\subsubsection{Rating Prediction}

As in the vanilla MF, the ratings can be modeled by user and item latent vectors using hierarchical embeddings.
Briefly, we only need to replace the latent matrices in Eq. (\ref{eq:mf-objective}) with the hierarchical embeddings.
However, the constraint of normalizing the connection matrices poses a challenge that makes it difficult to seamlessly apply a typical gradient-based optimizer.
To overcome this constraint, we instead apply a row-wise softmax function to the connection matrices.
Finally, we introduce HMF for rating prediction and the corresponding objective function is
\begin{align}
    \min_{\substack{\tilde{\boldsymbol{U}}^{(1)}, \cdots, \tilde{\boldsymbol{U}}^{(p)}, \boldsymbol{U}^{(p)}, \\ \tilde{\boldsymbol{V}}^{(1)}, \cdots, \tilde{\boldsymbol{V}}^{(q)}, \boldsymbol{V}^{(q)}}} \sum_{(i, j) \in \boldsymbol{\Omega}} \left( \boldsymbol{X}_{ij} - \boldsymbol{U}_i^{(0)T} \boldsymbol{V}_j^{(0)} \right)^2 + \lambda_{\boldsymbol{\Theta}} ||\boldsymbol{\Theta}||_2^2
\end{align}
where
\begin{align}
    \boldsymbol{U}^{(0)} &= \tilde{\tilde{\boldsymbol{U}}}^{(1)} \tilde{\tilde{\boldsymbol{U}}}^{(2)} \cdots \tilde{\tilde{\boldsymbol{U}}}^{(p)} \boldsymbol{U}^{(p)}, \boldsymbol{V}^{(0)} = \tilde{\tilde{\boldsymbol{V}}}^{(1)} \tilde{\tilde{\boldsymbol{V}}}^{(2)} \cdots \tilde{\tilde{\boldsymbol{V}}}^{(q)} \boldsymbol{V}^{(q)} \label{eq:constStart} \\
    \tilde{\tilde{\boldsymbol{U}}}^{(l)} &= \mathrm{softmax}(\tilde{\boldsymbol{U}}^{(l)}), \forall{l} \in \{1, \dots, p\} \\
    \tilde{\tilde{\boldsymbol{V}}}^{(l)} &= \mathrm{softmax}(\tilde{\boldsymbol{V}}^{(l)}), \forall{l} \in \{1, \dots, q\}.\label{eq:constEnd}
\end{align}

\subsubsection{Ranking Prediction}

HMF can be easily applied to ranking prediction algorithms that are optimized based on the gradient descent algorithm.
Here, we consider BPR-HMF, which is an extension of BPR to the HMF scheme.
The objective function is derived by replacing the MF terms in the BPR objective function (\ref{eq:bpr-objective}), as follows:
\begin{align}
    \min_{\substack{\tilde{\boldsymbol{U}}^{(1)}, \cdots, \tilde{\boldsymbol{U}}^{(p)}, \boldsymbol{U}^{(p)}, \\ \tilde{\boldsymbol{V}}^{(1)}, \cdots, \tilde{\boldsymbol{V}}^{(q)}, \boldsymbol{V}^{(q)}}} - \sum_{\substack{(i, j) \in \boldsymbol{\Omega},\\ (i, k) \notin \boldsymbol{\Omega}}}
    \ln \sigma(\boldsymbol{U}_i^{(0)T} &\boldsymbol{V}_j^{(0)} - \boldsymbol{U}_i^{(0)T} \boldsymbol{V}_k^{(0)}) + \lambda_{\boldsymbol{\Theta}} ||\boldsymbol{\Theta}||_2^2
\end{align}
where $\boldsymbol{U}^{(0)}$ and $\boldsymbol{V}^{(0)}$ follow Eq. (\ref{eq:constStart})--(\ref{eq:constEnd}).

{
\subsection{Interpretation}

The background of HMF interpretability can be illustrated by an interaction $\boldsymbol{U}^{(0)T}_i \boldsymbol{V}^{(0)}_j$ between a user $i$ and an item $j$.
From Eq. (6) and (7), the interaction can be transformed as
\begin{align}
    \boldsymbol{U}^{(0)T}_i \boldsymbol{V}^{(0)}_j &= (\tilde{\tilde{\boldsymbol{U}}}^{(1)T}_i \tilde{\tilde{\boldsymbol{U}}}^{(1)})^T \tilde{\tilde{\boldsymbol{V}}}^{(1)T}_j \tilde{\tilde{\boldsymbol{V}}}^{(1)} \\
    &= \sum_{s} \sum_{t} \tilde{\tilde{\boldsymbol{U}}}^{(1)}_{is} \tilde{\tilde{\boldsymbol{V}}}^{(1)}_{jt} \tilde{\tilde{\boldsymbol{U}}}^{(1)T}_s \tilde{\tilde{\boldsymbol{V}}}^{(1)}_t.
\end{align}
Significantly, the interactions between users and items are modeled by the inner product of the embeddings for the user and item clusters.
In addition, the selection of which cluster to use for prediction is determined by a coefficient ($\tilde{\tilde{\boldsymbol{U}}}^{(1)}_{is} \tilde{\tilde{\boldsymbol{V}}}^{(1)}_{jt}$) that indicates the strength of the relationship between the cluster and the entity (i.e., user or item).
In this way, since the relationship between clusters and entities is consistent with modeling interactions, it is natural that clusters provide HMF interpretability.
}
\section{Experimental Setup}

\subsection{Datasets}

\begin{table}[t]
    \caption{Dataset statistics after filtering.}
    \fontsize{7.5pt}{7.5pt}\selectfont
    \label{tab:dataset-statistics}
    \centering
    \begin{tabular}{lccc} \hline
        \toprule
        & \multirow{2}{*}{User / Item \#} & Interaction \# & \multirow{2}{*}{Density} \\ 
        &  & (training / validation / test) &  \\ \midrule
        ML-100K     & 625 / 1,561 & 64,000~/~2,775~/~1,932 & 0.0704 \\
        ML-1M       & 4,463 / 3,594 & 640,133~/~33,344~/~68,816 & 0.0463 \\
        Ciao        & 1,786 / 9,004 & 23,081~/~709~/~211 & 0.0015 \\
        DIGINETICA  & 24,933 / 54,859 & 180,894~/~2,324~/~1,398 & 0.0004 \\
        \bottomrule
    \end{tabular}
\end{table}

For the evaluation, we performed rating and ranking prediction tasks utilizing four datasets: ML-100K, ML-1M, Ciao, and DIGINETICA.
ML-100K and ML-1M~\cite{2015dec_f.harper} are datasets from the movie domain that contain a total of 100,000 and 1,000,209 five-star ratings given to movies by users, respectively.
As a product review dataset, we also used the released part\footnote{\url{https://www.cse.msu.edu/~tangjili/datasetcode/truststudy.htm}} of a dataset called Ciao~\cite{2012feb_j.tang}, which contains 36,065 ratings given to products by users.
DIGINETICA\footnote{\url{https://competitions.codalab.org/competitions/11161}} which contains user sessions in the e-commerce website, is used for the ranking prediction task.
We used only item view data (\verb|train-item-views.csv|) from January 1 to June 1 for this experiment. Some sessions were missing user IDs or had multiple user IDs. Therefore, the last user ID in each session was used as the user ID for that session, and sessions with no IDs were deleted.
Finally, the view between user and item was treated as implicit feedback.

Each dataset was divided into three subsets: training, validation, and testing.
Temporal global split~\cite{2020nov_z.meng} was used as the splitting strategy for evaluation in a realistic prediction task.
For ML-100K, ML-1M, and Ciao, all the interactions were sorted by timestamp and split into training (80\%) and testing (20\%) subsets.
Furthermore, the last 20\% of the training subset was used for validation.
Note that, unlike previous studies, users and items with a few ratings were not deleted.
In DIGINETICA, interactions were divided by timestamp: January to March into the training subset, April into the validation subset, and May and June 1 into the testing subset.
Users with less than five views in the training subset were removed from all subsets.
None of the methods, including the baselines, considered cold-start users or items.
Therefore, users and items that were not included in the training subset were removed from the validation and test evaluations.
Table \ref{tab:dataset-statistics} lists the statistics of the dataset used after these splits and filters.

\subsection{Baseline Methods}

\begin{table}[t]
    \caption{Hyperparameter settings for the models in this study.}
    \label{tab:hyperparameter-setting}
    \centering
    \fontsize{7.5pt}{7.5pt}\selectfont
    \begin{tabular}{lll} \hline
        \toprule
        Model & Hyperparameter & Range \\ \midrule
        All & Embed. size $d$ & $20$ \\ \midrule
        MF, HMF, & Weight decay & $10^{-2}, 10^{-3}, 10^{-4}, 10^{-5}, 0$ \\
        BPR-MF, NeuMF, & Learning rate & $10^{-2}, 10^{-3}, 10^{-4}$ \\
        ProtoMF, BPR-HMF & Batch size & $1024$ \\ \midrule
        MF, HMF & \# of epoch & 512 \\ \midrule
        BPR-MF, NeuMF, & \# of epoch & 128 \\
        ProtoMF, BPR-HMF & & \\ \midrule
        IHSR, HMF & \# of user clusters & $200, 400, 600, 800, 1000$ \\
        & \# of item clusters & $100, 200, 300, 400, 500$ \\ \midrule
        ProtoMF, BPR-HMF & \# of user clusters & $\lfloor24933/512\rfloor, \lfloor24933/128\rfloor,$ \\
        & & $\lfloor24933/32\rfloor$ \\
        & \# of item clusters & $\lfloor54859/512\rfloor, \lfloor54859/128\rfloor,$ \\
        & & $\lfloor54859/32\rfloor$ \\ \midrule
        IHSR & Reg. param. $\lambda$ & $0, 10^{-4}, 10^{-3}, 10^{-2}, 10^{-1}, 1, 10$ \\
        & Max. \# of iter. & $64$ \\ \midrule
        NeuMF & Layer sizes & $\{40, 20, 10\}$ \\ \midrule
        eTREE & \# of item clusters & $\{10\}, \{25, 5\}, \{50, 10, 3\}$ \\
        & Reg. param. $\lambda, \mu$ & $10^{-3}, 0.5, 1, 5, 10, 15, 20$ \\
        & Reg. param. $\eta$ & 1000 \\ \midrule
        ProtoMF & Tuning param. $\lambda_*$ & $10^{-3}, 10^{-2}, 10^{-1}, 1, 10$ \\
        \bottomrule
    \end{tabular}
\end{table}

The proposed methods were compared with several vanilla and state-of-the-art hierarchical MF methods to demonstrate their effectiveness.
In rating prediction (i.e., explicit feedback) tasks, we used three baselines: MF~\cite{2009aug_y.koren}, IHSR~\cite{2015jul_s.wang,2018jun_s.wang}, and eTREE~\cite{2021may_f.almutairi} for the proposed HMF.
In ranking prediction (i.e., implicit feedback) tasks, we compared BPR-HMF with three baselines: BPR-MF~\cite{2009jun.s.rendle}, NeuMF~\cite{2017apr_x.he}, ProtoMF~\cite{2022nov_a.melchiorre}.
All methods had certain hyperparameters.
The number of embedding dimensions for all methods was fixed at 20, and the other parameters were tuned in the validation set using a grid search.
Table \ref{tab:hyperparameter-setting} lists the details of the grid search range setting for each method.
For the sake of fairness, all the methods were compared with a depth one, except for eTREE, for which the hyperparameter settings are publicly available.
For the gradient descent methods, AdamW~\cite{2019jan_i.loshchilov} was used as the optimizer and weight decay was used instead of the model parameter regularization term.

\subsection{Evaluation Metrics}

For the rating prediction tasks, we applied RMSE to calculate differences between true rating $\boldsymbol{X}_{ij}$ and predicted rating $\hat{\boldsymbol{X}}_{ij}$ for observed interaction set in each subset, which is the standard for evaluating regression tasks. To evaluate the ranking prediction task, we prepared 100 item candidates, of which one was a positive-interacting item and 99 were negative-interacting items.
That is, for a given observed interaction (user $i$ views item $j$) in the evaluation subset, we randomly sampled 99 items that had not been viewed by user $i$ (i.e., negative sampling).
The 100 items were then ranked by a trained model and scored using HitRatio@$10$ and MRR@$10$.
To reduce the experimental duration, the training was terminated if the learning results did not improve for five consecutive epochs (or iterations) on the validation set.
We trained all the methods with five different seeds, and for each method, the hyperparameter set with the highest average results based on the RMSE and HitRatio of the validation subset was selected.
The average of the five evaluations of the testing subset was then reported.
\section{Results}

\subsection{Accuracy Comparison}

\begin{table}[t]
    \tabcolsep = 3pt
    \centering
    \caption{Evaluation results for the proposed and baseline methods. The sign $\dag$ indicates a significant difference over the proposed method (i.e., HMF or BPR-HMF) using Tukey's HSD test. All the runtimes were measured on the same machine (CPU: Intel Xeon Gold 6132, GPU: GeForce RTX 3080 Ti), with IHSR and eTREE implemented using NumPy and the other methods using JAX.}
    \label{tab:evaluation-results}
    \footnotesize
    \subtable[Rating prediction]{
        \begin{tabular}{lllrr}
            \toprule
            \multirow{2}{*}{Dataset} & \multirow{2}{*}{Method} & Accuracy $\pm$ std. & \multicolumn{2}{l}{Time on CPU / GPU [sec.]} \\ \cmidrule(lr){3-3}\cmidrule(lr){4-5}
            & & RMSE & Train & Inference \\
            \midrule
            \multirow{4}{*}{ML-100K} & MF    & 1.113 $\pm$ 0.013$\dag$ & 6.3 / \hspace{8pt}6.3 & 0.006 / 0.009 \\
                                     & IHSR  & 1.080 $\pm$ 0.002$\dag$ & 0.6 / \hspace{15.5pt}- & 0.009 / \hspace{15.5pt}- \\
                                     & eTREE & 1.082 $\pm$ 0.006$\dag$ & 182.6 / \hspace{15.5pt}- & 0.002 / \hspace{15.5pt}- \\
                                     & HMF   & \textbf{1.066} $\pm$ 0.002 & 82.7 / \hspace{4pt}10.7 & 0.024 / 0.014 \\
            \midrule
            \multirow{4}{*}{ML-1M} & MF    & \textbf{0.913} $\pm$ 0.002$\dag$ & 237.7 / 172.1 & 0.009 / 0.010 \\
                                   & IHSR  & 0.951 $\pm$ 0.000$\dag$ & 130.7 / \hspace{15.5pt}- & 0.227 / \hspace{15.5pt}- \\
                                   & eTREE & 0.922 $\pm$ 0.003 & 3054.5 / \hspace{15.5pt}- & 0.014 / \hspace{15.5pt}- \\
                                   & HMF   & 0.920 $\pm$ 0.003 & 209.1 / \hspace{8pt}7.0 & 0.557 / 0.016 \\
            \midrule
            \multirow{4}{*}{Ciao} & MF    & 2.306 $\pm$ 0.017$\dag$ & 21.9 / \hspace{4pt}15.6 & 0.007 / 0.010 \\
                                  & IHSR  & 1.122 $\pm$ 0.001$\dag$ & 84.9 / \hspace{15.5pt}- & 0.002 / \hspace{15.5pt}- \\
                                  & eTREE & 1.254 $\pm$ 0.045$\dag$ & 105.1 / \hspace{15.5pt}- & 0.020 / \hspace{15.5pt}- \\
                                  & HMF   & \textbf{0.939} $\pm$ 0.001 & 331.2 / \hspace{8pt}9.9 & 0.061 / 0.014 \\
            \bottomrule
        \end{tabular}
    }
    \subtable[Ranking prediction@10 (Dataset: DIGINETICA)]{
        \begin{tabular}{lllrr}
            \toprule
            \multirow{2}{*}{Method} & \multicolumn{2}{l}{Accuracy $\pm$ std.} & \multicolumn{2}{l}{Time on CPU / GPU [sec.]} \\ \cmidrule(lr){2-3}\cmidrule(lr){4-5}
            & HitRatio & MRR & Train & Inference \\
            \midrule
            BPR-MF  & 0.314 $\pm$ 0.019 & \textbf{0.153} $\pm$ 0.013$\dag$ & 160.4 / \hspace{4pt}62.7 & 0.025 / 0.016 \\
            NeuMF   & 0.307 $\pm$ 0.009 & 0.150 $\pm$ 0.005$\dag$ & 1094.2 / 290.2 & 0.142 / 0.030 \\
            ProtoMF & 0.277 $\pm$ 0.027$\dag$ & 0.122 $\pm$ 0.011 & 891.6 / \hspace{4pt}88.5 & 1.188 / 0.038 \\
            BPR-HMF & \textbf{0.323} $\pm$ 0.004 & 0.126 $\pm$ 0.003 & 1674.5 / \hspace{4pt}70.2 & 0.162 / 0.022 \\
            \bottomrule
        \end{tabular}
    }
\end{table}

Table \ref{tab:evaluation-results} presents the accuracy results of the proposed and the baseline methods for the testing subsets.
{In addition, the averages of the training and inference run times in the test are given for reference.}
On the small dataset ML-100K, the hierarchical methods IHSR, eTREE, and HMF showed superior accuracy, with the RMSE of HMF being 0.014 points lower than that of the second best, IHSR.
For the relatively large dataset ML-1M, HMF had the best accuracy among the hierarchical methods, whereas vanilla MF had the best accuracy overall.
This slight degradation may be due to the insufficient tuning of the hyperparameter set for ML-1M, which has a large number of users and items.
Specifying larger user or item clusters may have allowed HMF to outperform or match vanilla MF.
For Ciao, a sparse dataset compared to the others, HMF showed a dramatic improvement over MF, which did not converge well.
This suggests that a strong assumption of representative points can facilitate solution searching and provide robustness for sparse datasets.
For ranking prediction and DIGINETICA, the proposed BPR-HMF method was the best in terms of HitRatio, but not MRR.
This is because HMF assumes groups of items, which may make it difficult to push the rank of a particular item higher within a group, even though it may be possible to push the rank of the group to the top.
{The proposed methods (namely, HMF and BPR-HMF) on GPU took 0.04 to 1.70 times the training time and 1.38 to 1.56 times the inference time compared to the classical methods (namely, MF and BPR-MF).
This indicates that incorporating hierarchical embeddings does not significantly increase computational time.
Although the training times of the proposed methods on CPU were up to 138 times longer than those of the baseline methods, using a GPU can reduce them to a more practical execution time.}

\subsection{Loss Convergence}

\begin{figure*}[t!]
    \centering
    \subfigure[ML-100K]{
        \includegraphics[width=5.85cm]{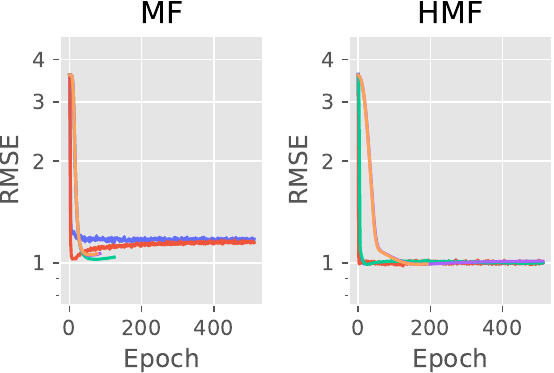}
    }
    \subfigure[ML-1M]{
        \includegraphics[width=5.85cm]{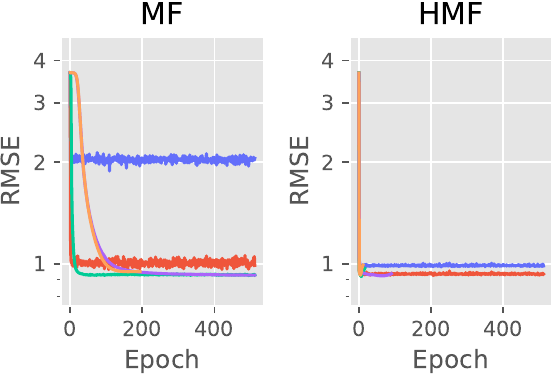}
    }
    \subfigure[Ciao]{
        \includegraphics[width=5.85cm]{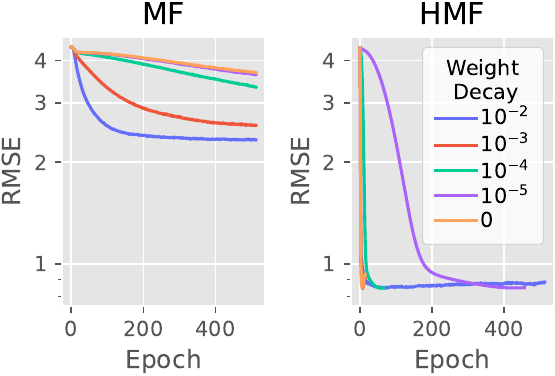}
    }
    \caption{MF and HMF losses (shown in RMSE) per epoch on the validation subsets. The best hyperparameter setting was selected for each weight decay, showing the change in its loss. Note that if the validation loss increased for five consecutive epochs, the optimization was terminated.}
    \label{fig:convergence-analysis}
\end{figure*}

To confirm the convergence of the HMF losses, which assumes clustering, we visualized the evolution of the losses for different hyperparameter settings.
We selected the best hyperparameter setting in the validation subset for each weight decay (which corresponded to the strength of the parameter regularization) and tracked the changes in its RMSE.
Fig. \ref{fig:convergence-analysis} shows the change in the RMSE of the validation subset per epoch for MF and HMF in the rating prediction tasks.
In ML-100K, MF tended to slightly overfit, even with strong regularization; however, HMF tended to converge near to one regardless of regularization.
This may be because of the clusters assumed by HMF, which may have had a regularization effect.
In other words, because each user or item is represented by a weighted average of the clusters, it is less likely to be assigned an inappropriate position in the latent space.
In ML-1M, MF required many epochs to converge, whereas HMF required approximately five epochs.
Therefore, even when using the same SGD-based optimization method, that is, AdamW, the architecture of the model can cause convergence.
For Ciao, which is considered difficult to learn owing to its high sparsity, MF did not converge well in any setting, whereas HMF converged to an RMSE of less than one in all setting.
This suggests that hierarchical embedding in HMF not only speeds up convergence but also increases the probability of convergence.

\subsection{Case Study}

\begin{table*}[t!]
    \vspace{-10pt}
    \centering
    \caption{A case study investigating the relationships between user and item clusters on ML-1M. For each target user (item) cluster, the inner product with all item (user) clusters is computed and sorted in descending order.}
    \label{tab:case-study}
    \fontsize{7pt}{7pt}\selectfont
    \begin{tabular}{cccc}
        \toprule
        \multicolumn{4}{c}{(a) Target: 0th User Cluster} \\ \cmidrule(lr){1-4}
        Item Cluster ID & Item Cluster Size & Item Cluster Titles & Inner Product \\
        \midrule
        125 & 8.91 & ``Bride of the Monster'', ``Godzilla (Gojira)'', ``Rocky Horror Picture Show, The'' & 25.19 \\
        384 & 10.55 & ``Braindead'', ``Bad Lieutenant'', ``Bride of the Monster'' & 25.02 \\
        187 & 9.84 & ``Godzilla (Gojira)'', ``Night of the Living Dead'', ``Pee-wee's Big Adventure'' & 24.80 \\
        $\dots$ & $\dots$ & $\dots$ & $\dots$ \\
        309 & 7.50 & ``Patriot, The'', ``Titanic'', ``Pretty Woman'' & -14.30 \\
        122 & 7.26 & ``Forrest Gump'', ``Patriot, The'', ``Shawshank Redemption, The'' & -16.26 \\
        114 & 7.32 & ``Titanic'', ``Lethal Weapon'', ``Independence Day (ID4)'' & -17.98 \\
        \toprule
        \multicolumn{4}{c}{(b) Target: 1st User Cluster} \\ \cmidrule(lr){1-4}
        Item Cluster ID & Item Cluster Size & Item Cluster Titles & Inner Product \\
        \midrule
        288 & 7.58 & ``Fargo'', ``Rushmore'', ``American Beauty'' & 19.98 \\
        75 & 8.05 & ``Austin Powers: The Spy Who Shagged Me'', ``Clockwork Orange, A'', ``Austin Powers: International Man of Mystery'' & 18.66 \\
        349 & 7.87 & ``Fear and Loathing in Las Vegas'', ``Pulp Fiction'', ``Star Wars: Episode V - The Empire Strikes Back'' & 17.15 \\
        $\dots$ & $\dots$ & $\dots$ & $\dots$ \\
        335 & 7.40 & ``Nil By Mouth'', ``For Love of the Game'', ``Girlfight'' & -13.91 \\
        416 & 7.89 & ``For Love of the Game'', ``Where the Heart Is'', ``Bed of Roses'' & -14.24 \\
        327 & 7.73 & ``Money Talks'', ``Son of Flubber'', ``Where the Heart Is'' & -14.65 \\
        \toprule
        \multicolumn{4}{c}{(c) Target: 374th Item Cluster} \\ \cmidrule(lr){1-4}
        User Cluster ID & User Cluster Size & User Cluster Genders & Inner Product \\
        \midrule
        35 & 5.24 & M, M, M, M, M, M, M, M, M, M & 23.07 \\
        232 & 5.19 & M, M, M, M, M, M, M, M, M, M & 22.53 \\
        404 & 7.28 & M, M, F, M, M, M, M, M, F, F & 20.68 \\
        $\dots$ & $\dots$ & $\dots$ & $\dots$ \\
        86 & 6.43 & M, M, F, F, F, M, F, M, M, M & -14.62 \\
        400 & 7.16 & F, M, M, F, F, F, M, M, F, F & -15.24 \\
        624 & 7.85 & F, M, M, F, M, M, M, M, M, M & -15.46 \\
        \toprule
        \multicolumn{4}{c}{(d) Target: 216th Item Cluster} \\ \cmidrule(lr){1-4}
        User Cluster ID & User Cluster Size & User Cluster Genders & Inner Product \\
        \midrule
        156 & 8.35 & F, M, F, F, M, F, M, M, F, M & 30.95 \\
        61 & 6.26 & F, F, F, M, M, F, F, M, M, F & 29.17 \\
        200 & 6.62 & F, F, F, F, M, M, F, M, F, F & 27.89 \\
        $\dots$ & $\dots$ & $\dots$ & $\dots$ \\
        94 & 9.28 & M, M, M, M, M, M, M, M, M, M & -20.89 \\
        764 & 7.54 & M, F, M, M, M, M, M, M, M, M & -23.41 \\
        347 & 5.91 & M, M, M, M, M, M, M, M, M, M & -23.45 \\
        \bottomrule
    \end{tabular}
\end{table*}

\begin{table}[t]
    \vspace{-10pt}
    \footnotesize
    \centering
    \caption{A case study investigating the relationships between users/items and user/item clusters on ML-1M. For each target user (item) cluster, the probability of connecting to all users (items) is sorted in descending order.}
    \label{tab:case-study-2}
    \subtable[1st User Cluster -- Users]{
        \begin{tabular}{cccc}
            \toprule
            User ID & Connect. Probab. & Gender & Occupation \\
            \midrule
            2915 & 0.048 & F & K-12 student \\
            4162 & 0.046 & M & academic/educator \\
            2960 & 0.024 & M & programmer \\
            773 & 0.024 & M & programmer \\
            3338 & 0.019 & M & writer \\
            $\dots$ & $\dots$ & $\dots$ & $\dots$ \\
            \bottomrule
        \end{tabular}
    }
    \subtable[374th Item Cluster -- Items]{
        \begin{tabular}{ccc}
            \toprule
            Item ID & Connect. Probab. & Item Title \\
            \midrule
            285 & 0.196 & Pulp Fiction \\
            251 & 0.196 & Star Wars: Episode IV - A New Hope \\
            1066 & 0.172 & Star Wars: Episode V - The Empire ... \\
            1080 & 0.146 & Star Wars: Episode VI - Return of ... \\
            3338 & 0.059 & For a Few Dollars More (1965) \\
            $\dots$ & $\dots$ & $\dots$ \\
            \bottomrule
        \end{tabular}
    }
\end{table}

In HMF, all users, items, and their clusters are projected onto the same latent space, allowing identifying whether user representatives (clusters) think highly about item categories (clusters) has significant benefits for enhancing service quality.
Herein, we present a case study of ML-1M for a movie service.
Table \ref{tab:case-study} shows the inner products of the zeroth and first user clusters and all the item clusters, as well as the inner products of the 374th and 216th item clusters and all the user clusters.
The inner product of the $i$-th user cluster and the $j$-th item cluster was computed as $\boldsymbol{U}_i^{(1)T} \boldsymbol{V}_j^{(1)}$.
The user/item cluster size was calculated based on the weights of the connection matrix with a user/item of 1; that is, the size of the $j$-th user cluster ID is $\sum_i \tilde{\tilde{\boldsymbol{U}}}_{ij}^{(1)}$.
{In addition, {\em item cluster titles} ({\em user cluster genders}) characterized the item (user) cluster and were the titles (genders) of items (users) in the neighborhood of the item (user) cluster in the latent space.
Table \ref{tab:case-study} (a) shows that the zeroth user cluster strongly prefers horror movies such as ``Bride of the Monster'' and ``Braindead,'' indicating that it captures similar movies through explicit feedback.
From \ref{tab:case-study} (c) and (d), it is possible that men prefer the 374th item cluster while the opposite trend was observed for the 216 th item cluster.
In addition to the relationship between user and item clusters, it is also interesting to observe the relationship between clusters and their entities, as shown in table \ref{tab:case-study-2}.
The connection matrix allows us to observe entities that strongly belong to a cluster, and it seems that users or items with close attributes were considered as the same cluster.
Other relationships, such as between users and item clusters, can also be observed, and by observing the various relationships between users, items, and clusters, HMF has interpretability that provides insight by summarizing the learning results.}

{

\begin{table}[t!]
    \vspace{-10pt}
    \centering
    \footnotesize
    \caption{Evaluation results for the proposed methods with different depths of hierarchical embeddings. The sign $\dag$ indicates a significant difference over the model with depth 1 using Tukey's HSD test.}
    \label{tab:evaluation-results-with-different-depth}
    \subtable[Rating Prediction]{
        \begin{tabular}{lllll}
            \toprule
            \multirow{2}{*}{Method} & \multirow{2}{*}{Depth $p, q$} & ML-100K & ML-1M & Ciao \\ \cmidrule(lr){3-3}\cmidrule(lr){4-4}\cmidrule(lr){5-5}
            & & RMSE $\pm$ std. & RMSE $\pm$ std. & RMSE $\pm$ std. \\
            \midrule
            \multirow{4}{*}{HMF} & 1 & 1.066 $\pm$ 0.002          & \textbf{0.920} $\pm$ 0.003 & 0.939 $\pm$ 0.001           \\
                                 & 2 & 1.070 $\pm$ 0.013          & 0.930 $\pm$ 0.003$\dag$    & 0.928 $\pm$ 0.005           \\
                                 & 3 & \textbf{1.061} $\pm$ 0.005 & 0.928 $\pm$ 0.004$\dag$    & \textbf{0.927} $\pm$ 0.012  \\
                                 & 4 & 1.067 $\pm$ 0.017          & 0.931 $\pm$ 0.002$\dag$    & 0.930 $\pm$ 0.011           \\
            \bottomrule
        \end{tabular}
    }
    \subtable[Ranking Prediction@10 (Dataset: DIGINETICA)]{
        \begin{tabular}{llll}
            \toprule
            Method & Depth $p, q$ & HitRatio $\pm$ std. & MRR $\pm$ std. \\
            \midrule
            \multirow{4}{*}{BPR-HMF} & 1 & 0.323 $\pm$ 0.004          & 0.126 $\pm$ 0.003 \\
                                     & 2 & 0.325 $\pm$ 0.003          & 0.125 $\pm$ 0.002 \\
                                     & 3 & \textbf{0.328} $\pm$ 0.010 & \textbf{0.127} $\pm$ 0.002 \\
                                     & 4 & 0.326 $\pm$ 0.008          & 0.126 $\pm$ 0.002 \\
            \bottomrule
        \end{tabular}
    }
\end{table}

\subsection{Influence of Hierarchy Depth}

So far, we have studied HMF with a hierarchical depth of one.
However, it is not clear how the depth affects the recommendation accuracy of HMF.
Therefore, this section compares the recommendation accuracy of the depth-one HMF with that of the deeper HMF.
Based on the number of user (item) clusters $m_1$ ($n_1$) selected in the depth-one HMF, we constructed a hierarchy in which the number of user (item) clusters at level $l$ is defined as $m_l = m_1 / 2^{l-1}$ ($n_l = n_1 / 2^{l-1}$), and we report the results of retuning hyperparameters without the number of clusters.
Table \ref{tab:evaluation-results-with-different-depth} presents the accuracy results of the proposed methods with depth $p = q = 1, 2, 3, 4$ for the testing subsets.
Except for ML-1M, where the proposed method's recommendation accuracy was inferior to that of the baseline mentioned in Section 5.1, no significant change in the recommendation accuracy of the proposed method is observed among different depths $p, q$.
Therefore, the reason why the proposed method outperforms the traditional methods in Table \ref{tab:evaluation-results} is not to increase the depth $p, q$ of the hierarchy, but to capture clusters.
The advantage of hierarchical embedding that we expect is interpretability, which can provide multiple granularities of interpretation results, rather than improved accuracy.
On the other hand, ML-1M tends to be less accurate with depth.
We have not yet identified the cause of this trend, but a detailed analysis of this trend may reveal the cause of the proposed method's inferiority to the baseline in ML-1M.
}
\section{Conclusion}

In this paper, we proposed HMF, which captures the hierarchical relationships between users and items for interpretable recommender systems.
It is assumed that the embedding vector of the user/item or cluster is the weighted average of the embedding vectors of the parent clusters in the hierarchical structure.
This simple formulation allowed us to tackle both MF and clustering with a single gradient method, and also facilitated the possibility of using it for recently developed MF methods based on the gradient method.
The experimental results on real datasets showed that our methods equaled or outperformed existing hierarchical and vanilla MF methods, demonstrating competitiveness and robustness in particularly sparse interactions.
By characterizing user and item clusters, we presented relationships between the clusters and provided an example of how we can interpret how HMF learns user and item interactions.
Nevertheless, the study has limitations we should solve in future work.
The most important one is that hierarchical embeddings may not be good at subtle ranking, therefore we need to control the strength of the clustering and be careful about the expressiveness of the embedding.

\section*{Acknowledgments}
This work was supported by JSPS KAKENHI, Grant Numbers JP21H03553 and JP22H03698.

\bibliographystyle{model5-names}
\bibliography{refs}

\begin{thebibliography}{29}
\expandafter\ifx\csname natexlab\endcsname\relax\def\natexlab#1{#1}\fi
\providecommand{\url}[1]{\texttt{#1}}
\providecommand{\href}[2]{#2}
\providecommand{\path}[1]{#1}
\providecommand{\DOIprefix}{doi:}
\providecommand{\ArXivprefix}{arXiv:}
\providecommand{\URLprefix}{URL: }
\providecommand{\Pubmedprefix}{pmid:}
\providecommand{\doi}[1]{\href{https://doi.org/#1}{\path{#1}}}
\providecommand{\Pubmed}[1]{\href{pmid:#1}{\path{#1}}}
\providecommand{\bibinfo}[2]{#2}
\ifx\xfnm\relax \def\xfnm[#1]{\unskip,\space#1}\fi
%Type = Inproceedings
\bibitem[{Abdollahi \& Nasraoui(2017)}]{2017aug_b.abdollahi}
\bibinfo{author}{Abdollahi, B.}, \& \bibinfo{author}{Nasraoui, O.} (\bibinfo{year}{2017}).
\newblock \bibinfo{title}{Using explainability for constrained matrix factorization}.
\newblock In {\it \bibinfo{booktitle}{Proc. of the Eleventh ACM Conf. on Recomm. Syst.}\/} (p. \bibinfo{pages}{79–83}).
%Type = Article
\bibitem[{Almutairi et~al.(2021)Almutairi, Wang, Wang, Zhao \& Sidiropoulos}]{2021may_f.almutairi}
\bibinfo{author}{Almutairi, F.~M.}, \bibinfo{author}{Wang, Y.}, \bibinfo{author}{Wang, D.}, \bibinfo{author}{Zhao, E.}, \& \bibinfo{author}{Sidiropoulos, N.~D.} (\bibinfo{year}{2021}).
\newblock \bibinfo{title}{etree: Learning tree-structured embeddings}.
\newblock {\it \bibinfo{journal}{Proc. of the AAAI Conf. on Artif. Intell.}\/},  {\it \bibinfo{volume}{35}\/}, \bibinfo{pages}{6609--6617}.
%Type = Inproceedings
\bibitem[{Bayer et~al.(2017)Bayer, He, Kanagal \& Rendle}]{2017apr_i.bayer}
\bibinfo{author}{Bayer, I.}, \bibinfo{author}{He, X.}, \bibinfo{author}{Kanagal, B.}, \& \bibinfo{author}{Rendle, S.} (\bibinfo{year}{2017}).
\newblock \bibinfo{title}{A generic coordinate descent framework for learning from implicit feedback}.
\newblock In {\it \bibinfo{booktitle}{Proc. of the 26th Int. Conf. on World Wide Web}\/} (p. \bibinfo{pages}{1341–1350}).
%Type = Article
\bibitem[{Chen et~al.(2021)Chen, Mensah, Ma, Wang \& Jiang}]{2021nov_y.chen}
\bibinfo{author}{Chen, Y.}, \bibinfo{author}{Mensah, S.}, \bibinfo{author}{Ma, F.}, \bibinfo{author}{Wang, H.}, \& \bibinfo{author}{Jiang, Z.} (\bibinfo{year}{2021}).
\newblock \bibinfo{title}{Collaborative filtering grounded on knowledge graphs}.
\newblock {\it \bibinfo{journal}{Pattern Recognit. Lett.}\/},  {\it \bibinfo{volume}{151}\/}, \bibinfo{pages}{55--61}.
%Type = Inproceedings
\bibitem[{Cheng et~al.(2019)Cheng, Shen, Huang \& Zhu}]{2019jul_w.cheng}
\bibinfo{author}{Cheng, W.}, \bibinfo{author}{Shen, Y.}, \bibinfo{author}{Huang, L.}, \& \bibinfo{author}{Zhu, Y.} (\bibinfo{year}{2019}).
\newblock \bibinfo{title}{Incorporating interpretability into latent factor models via fast influence analysis}.
\newblock In {\it \bibinfo{booktitle}{Proc. of the 25th ACM SIGKDD Int. Conf. on Knowl. Discov. \& Data Min.}\/} (p. \bibinfo{pages}{885–893}).
%Type = Article
\bibitem[{Harper \& Konstan(2015)}]{2015dec_f.harper}
\bibinfo{author}{Harper, F.~M.}, \& \bibinfo{author}{Konstan, J.~A.} (\bibinfo{year}{2015}).
\newblock \bibinfo{title}{The movielens datasets: History and context}.
\newblock {\it \bibinfo{journal}{ACM Trans. on Interact. Intell. Syst.}\/},  {\it \bibinfo{volume}{5}\/}.
%Type = Inproceedings
\bibitem[{He et~al.(2017)He, Liao, Zhang, Nie, Hu \& Chua}]{2017apr_x.he}
\bibinfo{author}{He, X.}, \bibinfo{author}{Liao, L.}, \bibinfo{author}{Zhang, H.}, \bibinfo{author}{Nie, L.}, \bibinfo{author}{Hu, X.}, \& \bibinfo{author}{Chua, T.-S.} (\bibinfo{year}{2017}).
\newblock \bibinfo{title}{Neural collaborative filtering}.
\newblock In {\it \bibinfo{booktitle}{Proc. of the 26th Int. Conf. on World Wide Web}\/} (p. \bibinfo{pages}{173–182}).
%Type = Inproceedings
\bibitem[{Hu et~al.(2008)Hu, Koren \& Volinsky}]{2008dec_y.hu}
\bibinfo{author}{Hu, Y.}, \bibinfo{author}{Koren, Y.}, \& \bibinfo{author}{Volinsky, C.} (\bibinfo{year}{2008}).
\newblock \bibinfo{title}{Collaborative filtering for implicit feedback datasets}.
\newblock In {\it \bibinfo{booktitle}{2008 Eighth IEEE Int. Conf. on Data Min.}\/} (pp. \bibinfo{pages}{263--272}).
%Type = Article
\bibitem[{Koren et~al.(2009)Koren, Bell \& Volinsky}]{2009aug_y.koren}
\bibinfo{author}{Koren, Y.}, \bibinfo{author}{Bell, R.}, \& \bibinfo{author}{Volinsky, C.} (\bibinfo{year}{2009}).
\newblock \bibinfo{title}{Matrix factorization techniques for recommender systems}.
\newblock {\it \bibinfo{journal}{Comput.}\/},  {\it \bibinfo{volume}{42}\/}, \bibinfo{pages}{30--37}.
%Type = Article
\bibitem[{Li et~al.(2019)Li, Liu, Qian, Mamoulis, Tu \& Cheung}]{2019may_h.li}
\bibinfo{author}{Li, H.}, \bibinfo{author}{Liu, Y.}, \bibinfo{author}{Qian, Y.}, \bibinfo{author}{Mamoulis, N.}, \bibinfo{author}{Tu, W.}, \& \bibinfo{author}{Cheung, D.~W.} (\bibinfo{year}{2019}).
\newblock \bibinfo{title}{Hhmf: hidden hierarchical matrix factorization for recommender systems}.
\newblock {\it \bibinfo{journal}{Data Min. and Knowl. Discov.}\/},  {\it \bibinfo{volume}{33}\/}, \bibinfo{pages}{1548–1582}.
%Type = Inproceedings
\bibitem[{Loshchilov \& Hutter(2019)}]{2019jan_i.loshchilov}
\bibinfo{author}{Loshchilov, I.}, \& \bibinfo{author}{Hutter, F.} (\bibinfo{year}{2019}).
\newblock \bibinfo{title}{Decoupled weight decay regularization}.
\newblock In {\it \bibinfo{booktitle}{Int. Conf. on Learn. Represent.}\/}.
%Type = Inproceedings
\bibitem[{Mashhoori \& Hashemi(2012)}]{2012xxx_ali.mashhoori}
\bibinfo{author}{Mashhoori, A.}, \& \bibinfo{author}{Hashemi, S.} (\bibinfo{year}{2012}).
\newblock \bibinfo{title}{Incorporating hierarchical information into the matrix factorization model for collaborative filtering}.
\newblock In {\it \bibinfo{booktitle}{Intell. Inf. and Database Syst.}\/} (pp. \bibinfo{pages}{504--513}).
%Type = Inproceedings
\bibitem[{McAuley \& Leskovec(2013)}]{2013oct_j.mcauley}
\bibinfo{author}{McAuley, J.}, \& \bibinfo{author}{Leskovec, J.} (\bibinfo{year}{2013}).
\newblock \bibinfo{title}{Hidden factors and hidden topics: Understanding rating dimensions with review text}.
\newblock In {\it \bibinfo{booktitle}{Proc. of the 7th ACM Conf. on Recomm. Syst.}\/} (p. \bibinfo{pages}{165–172}).
%Type = Inproceedings
\bibitem[{Melchiorre et~al.(2022)Melchiorre, Rekabsaz, Ganh\"{o}r \& Schedl}]{2022nov_a.melchiorre}
\bibinfo{author}{Melchiorre, A.~B.}, \bibinfo{author}{Rekabsaz, N.}, \bibinfo{author}{Ganh\"{o}r, C.}, \& \bibinfo{author}{Schedl, M.} (\bibinfo{year}{2022}).
\newblock \bibinfo{title}{Protomf: Prototype-based matrix factorization for effective and explainable recommendations}.
\newblock In {\it \bibinfo{booktitle}{Proc. of the 16th ACM Conf. on Recomm. Syst.}\/} (p. \bibinfo{pages}{246–256}).
%Type = Inproceedings
\bibitem[{Meng et~al.(2020)Meng, McCreadie, Macdonald \& Ounis}]{2020nov_z.meng}
\bibinfo{author}{Meng, Z.}, \bibinfo{author}{McCreadie, R.}, \bibinfo{author}{Macdonald, C.}, \& \bibinfo{author}{Ounis, I.} (\bibinfo{year}{2020}).
\newblock \bibinfo{title}{Exploring data splitting strategies for the evaluation of recommendation models}.
\newblock In {\it \bibinfo{booktitle}{Proc. of the 14th ACM Conf. on Recomm. Syst.}\/} (p. \bibinfo{pages}{681–686}).
%Type = Inproceedings
\bibitem[{Menon et~al.(2011)Menon, Chitrapura, Garg, Agarwal \& Kota}]{2011aug_a.menon}
\bibinfo{author}{Menon, A.~K.}, \bibinfo{author}{Chitrapura, K.-P.}, \bibinfo{author}{Garg, S.}, \bibinfo{author}{Agarwal, D.}, \& \bibinfo{author}{Kota, N.} (\bibinfo{year}{2011}).
\newblock \bibinfo{title}{Response prediction using collaborative filtering with hierarchies and side-information}.
\newblock In {\it \bibinfo{booktitle}{Proc. of the 17th ACM SIGKDD Int. Conf. on Knowl. Discov. and Data Min.}\/} (p. \bibinfo{pages}{141–149}).
%Type = Article
\bibitem[{Montavon et~al.(2018)Montavon, Samek \& Müller}]{2018feb_g.montavon}
\bibinfo{author}{Montavon, G.}, \bibinfo{author}{Samek, W.}, \& \bibinfo{author}{Müller, K.-R.} (\bibinfo{year}{2018}).
\newblock \bibinfo{title}{Methods for interpreting and understanding deep neural networks}.
\newblock {\it \bibinfo{journal}{Digit. Signal Process.}\/},  {\it \bibinfo{volume}{73}\/}, \bibinfo{pages}{1--15}.
%Type = Inproceedings
\bibitem[{Ning \& Karypis(2011)}]{2011dec_x.ning}
\bibinfo{author}{Ning, X.}, \& \bibinfo{author}{Karypis, G.} (\bibinfo{year}{2011}).
\newblock \bibinfo{title}{Slim: Sparse linear methods for top-n recommender systems}.
\newblock In {\it \bibinfo{booktitle}{2011 IEEE 11th Int. Conf. on Data Min.}\/} (pp. \bibinfo{pages}{497--506}).
%Type = Inproceedings
\bibitem[{Oh et~al.(2001)Oh, Honda \& Ichihashi}]{2001jul_c.oh}
\bibinfo{author}{Oh, C.-H.}, \bibinfo{author}{Honda, K.}, \& \bibinfo{author}{Ichihashi, H.} (\bibinfo{year}{2001}).
\newblock \bibinfo{title}{Fuzzy clustering for categorical multivariate data}.
\newblock In {\it \bibinfo{booktitle}{Proc. Joint 9th IFSA World Congr. and 20th NAFIPS Int. Conf.}\/} (pp. \bibinfo{pages}{2154--2159}).
\newblock volume~\bibinfo{volume}{4}.
%Type = Inproceedings
\bibitem[{Pan et~al.(2008)Pan, Zhou, Cao, Liu, Lukose, Scholz \& Yang}]{2008dec_r.pan}
\bibinfo{author}{Pan, R.}, \bibinfo{author}{Zhou, Y.}, \bibinfo{author}{Cao, B.}, \bibinfo{author}{Liu, N.~N.}, \bibinfo{author}{Lukose, R.}, \bibinfo{author}{Scholz, M.}, \& \bibinfo{author}{Yang, Q.} (\bibinfo{year}{2008}).
\newblock \bibinfo{title}{One-class collaborative filtering}.
\newblock In {\it \bibinfo{booktitle}{2008 Eighth IEEE Int. Conf. on Data Min.}\/} (pp. \bibinfo{pages}{502--511}).
%Type = Inbook
\bibitem[{Rendle(2022)}]{2022xxx_f.ricci}
\bibinfo{author}{Rendle, S.} (\bibinfo{year}{2022}).
\newblock \bibinfo{title}{Item recommendation from implicit feedback}.
\newblock In \bibinfo{editor}{F.~Ricci}, \bibinfo{editor}{L.~Rokach}, \& \bibinfo{editor}{B.~Shapira} (Eds.), {\it \bibinfo{booktitle}{Recomm. Syst. Handb.}\/} (pp. \bibinfo{pages}{143--171}).
%Type = Inproceedings
\bibitem[{Rendle et~al.(2009)Rendle, Freudenthaler, Gantner \& Schmidt-Thieme}]{2009jun.s.rendle}
\bibinfo{author}{Rendle, S.}, \bibinfo{author}{Freudenthaler, C.}, \bibinfo{author}{Gantner, Z.}, \& \bibinfo{author}{Schmidt-Thieme, L.} (\bibinfo{year}{2009}).
\newblock \bibinfo{title}{Bpr: Bayesian personalized ranking from implicit feedback}.
\newblock In {\it \bibinfo{booktitle}{Proc. of the Twenty-Fifth Conf. on Uncertain. in Artif. Intell.}\/} (p. \bibinfo{pages}{452–461}).
%Type = Inproceedings
\bibitem[{Tang et~al.(2012)Tang, Gao \& Liu}]{2012feb_j.tang}
\bibinfo{author}{Tang, J.}, \bibinfo{author}{Gao, H.}, \& \bibinfo{author}{Liu, H.} (\bibinfo{year}{2012}).
\newblock \bibinfo{title}{Mtrust: Discerning multi-faceted trust in a connected world}.
\newblock In {\it \bibinfo{booktitle}{Proc. of the Fifth ACM Int. Conf. on Web Search and Data Min.}\/} (p. \bibinfo{pages}{93–102}).
%Type = Inproceedings
\bibitem[{Wang et~al.(2015)Wang, Tang, Wang \& Liu}]{2015jul_s.wang}
\bibinfo{author}{Wang, S.}, \bibinfo{author}{Tang, J.}, \bibinfo{author}{Wang, Y.}, \& \bibinfo{author}{Liu, H.} (\bibinfo{year}{2015}).
\newblock \bibinfo{title}{Exploring implicit hierarchical structures for recommender systems}.
\newblock In {\it \bibinfo{booktitle}{Proc. of the 24th Int. Conf. on Artif. Intell.}\/} (p. \bibinfo{pages}{1813–1819}).
%Type = Article
\bibitem[{Wang et~al.(2018)Wang, Tang, Wang \& Liu}]{2018jun_s.wang}
\bibinfo{author}{Wang, S.}, \bibinfo{author}{Tang, J.}, \bibinfo{author}{Wang, Y.}, \& \bibinfo{author}{Liu, H.} (\bibinfo{year}{2018}).
\newblock \bibinfo{title}{Exploring hierarchical structures for recommender systems}.
\newblock {\it \bibinfo{journal}{IEEE Trans. on Knowl. and Data Eng.}\/},  {\it \bibinfo{volume}{30}\/}, \bibinfo{pages}{1022--1035}.
%Type = Inproceedings
\bibitem[{Wang et~al.(2014)Wang, Pan \& Xu}]{2014nov_x.wang}
\bibinfo{author}{Wang, X.}, \bibinfo{author}{Pan, W.}, \& \bibinfo{author}{Xu, C.} (\bibinfo{year}{2014}).
\newblock \bibinfo{title}{Hgmf: Hierarchical group matrix factorization for collaborative recommendation}.
\newblock In {\it \bibinfo{booktitle}{Proc. of the 23rd ACM Int. Conf. on Inf. and Knowl. Manag.}\/} (p. \bibinfo{pages}{769–778}).
%Type = Inproceedings
\bibitem[{Xian et~al.(2021)Xian, Zhao, Li, Chan, Kan, Ma, Dong, Faloutsos, Karypis, Muthukrishnan \& Zhang}]{2021sep_y.xian}
\bibinfo{author}{Xian, Y.}, \bibinfo{author}{Zhao, T.}, \bibinfo{author}{Li, J.}, \bibinfo{author}{Chan, J.}, \bibinfo{author}{Kan, A.}, \bibinfo{author}{Ma, J.}, \bibinfo{author}{Dong, X.~L.}, \bibinfo{author}{Faloutsos, C.}, \bibinfo{author}{Karypis, G.}, \bibinfo{author}{Muthukrishnan, S.}, \& \bibinfo{author}{Zhang, Y.} (\bibinfo{year}{2021}).
\newblock \bibinfo{title}{Ex3: Explainable attribute-aware item-set recommendations}.
\newblock In {\it \bibinfo{booktitle}{Proc. of the 15th ACM Conf. on Recomm. Syst.}\/} (p. \bibinfo{pages}{484–494}).
%Type = Inproceedings
\bibitem[{Xue et~al.(2017)Xue, Dai, Zhang, Huang \& Chen}]{2017aug_h.xue}
\bibinfo{author}{Xue, H.-J.}, \bibinfo{author}{Dai, X.-Y.}, \bibinfo{author}{Zhang, J.}, \bibinfo{author}{Huang, S.}, \& \bibinfo{author}{Chen, J.} (\bibinfo{year}{2017}).
\newblock \bibinfo{title}{Deep matrix factorization models for recommender systems}.
\newblock In {\it \bibinfo{booktitle}{Proc. of the 26th Int. Joint Conf. on Artif. Intell.}\/} (p. \bibinfo{pages}{3203–3209}).
%Type = Article
\bibitem[{Zhang \& Chen(2020)}]{2020xxx_y.zhang}
\bibinfo{author}{Zhang, Y.}, \& \bibinfo{author}{Chen, X.} (\bibinfo{year}{2020}).
\newblock \bibinfo{title}{Explainable recommendation: A survey and new perspectives}.
\newblock {\it \bibinfo{journal}{Found. and Trends® in Inf. Retr.}\/},  {\it \bibinfo{volume}{14}\/}, \bibinfo{pages}{1--101}.

\end{thebibliography}

% \section*{Supplementary Material}

% Supplementary material that may be helpful in the review process should
% be prepared and provided as a separate electronic file. That file can
% then be transformed into PDF format and submitted along with the
% manuscript and graphic files to the appropriate editorial office.

\end{document}